\documentclass[11pt,reqno]{article}

\usepackage{color}
\usepackage{dcolumn}
\usepackage{bm}
\usepackage{amsmath}
\usepackage{amsfonts}
\usepackage{amssymb}
\usepackage{hyperref}
\usepackage{latexsym}
\usepackage{epsf}

\allowdisplaybreaks 

\renewcommand{\baselinestretch}{1.18}
\newcommand{\be}{\begin{equation}}
\newcommand{\ee}{\end{equation}}
\newcommand{\bea}{\begin{eqnarray}}
\newcommand{\eea}{\end{eqnarray}}

\let\a=\alpha     
     \let\th=\theta   \let\l=\lambda
\let\m=\mu    \let\n=\nu          \let\r=\rho 
\let\s=\sigma      
    
 \let\D=\Delta   \let\L=\Lambda 
         
  \let\eps=\epsilon

\let\==\equiv

\def\EE{{\cal E}}

\newcommand{\Tr}{{\rm Tr}}

\newcommand{\cO}{\mathcal{O}}
\newcommand{\cR}{\mathcal{R}}
\newcommand{\cS}{\mathcal{S}}

\newcommand{\half}{\tfrac{1}{2}}

\newcommand{\gb}{\bar{g}}

\newcommand{\p}{\partial}

\newcommand{\Db}{\bar{D}}

\newcommand{\Rb}{\bar{R}}

\begin{document}

\thispagestyle{empty}
\begin{flushright} \small
PI-QG-115; ITP--UU--09/03; SPIN--09/03; IPhT-T09/012
\end{flushright}
\bigskip

\begin{center}
 {\LARGE\bfseries Asymptotic Safety in Higher-Derivative Gravity}
\\[10mm]
Dario Benedetti$^1$, Pedro F. Machado$^2$ and Frank Saueressig$^{3}$ \\[3mm]
$^1${\small\slshape
Perimeter Institute for Theoretical Physics \\
31 Caroline St.\ N, N2L 2Y5, Waterloo ON, Canada \\
{\upshape\ttfamily dbenedetti@perimeterinstitute.ca} } \\[3mm]
$^2${\small\slshape
Institute for Theoretical Physics \emph{and} Spinoza Institute \\
Utrecht University, 3508 TD Utrecht, The Netherlands \\
{\upshape\ttfamily P.Machado@phys.uu.nl} }
\\[3mm]
$^3${\small\slshape
Institut de Physique Th\'eorique, CEA Saclay \\
91191 Gif-Sur-Yvette Cedex, France \\
{\upshape\ttfamily Frank.Saueressig@cea.fr} }\\

\end{center}
\vspace{5mm}

\hrule\bigskip

\begin{abstract}
We study the non-perturbative renormalization group flow of higher-derivative gravity employing functional renormalization group techniques. The non-perturbative contributions to the $\beta$-functions shift the known perturbative ultraviolet fixed point into a non-trivial fixed point with three UV-attractive and one UV-repulsive eigendirections, consistent with the asymptotic safety conjecture of gravity. The implication of this transition on the unitarity problem, typically haunting higher-derivative gravity theories, is discussed.


\end{abstract}

\vspace{5mm}
\hrule\bigskip
\vspace{5mm}


Among the many approaches to quantum gravity, a special place is occupied by higher-derivative gravity, which, besides the Einstein-Hilbert term, also includes fourth-order operators in the action. Indeed, the higher-derivative propagators soften the divergences encountered in the perturbative quantization, rendering the theory perturbatively renormalizable \cite{Stelle:1976gc} and asymptotically free at the one-loop level 
\cite{Julve:1978xn,Fradkin:1981hx,Avramidi:1985ki,deBerredoPeixoto:2004if,Codello:2006in}. 
Unfortunately, the extra terms responsible for the improved UV behavior also induce massive negative norm states \cite{Stelle:1977ry}, so-called ``poltergeists'', which led to the belief that the theory is not unitary. Several arguments suggest that this shortcoming can be cured by quantum effects \cite{Julve:1978xn,Salam:1978fd}, but the lack of non-perturbative methods has made it hard to substantiate such claims.

Recently, the question of renormalizability has received renewed attention due to mounting evidence 
in favor of the non-perturbative renormalizability, or \emph{asymptotic safety} (AS), of gravity \cite{Weinberg:1980gg,Niedermaier,Percacci:2007sz,Reuter:2007rv}.
In this scenario, the ultraviolet (UV) behavior 
of the theory is controlled by a non-Gaussian fixed point (NGFP) of the renormalization group flow, with a finite number of UV-attractive (relevant) directions. This fixed point is supposed to provide a continuum limit which is predictive and safe from divergences.
While initially found only for gravity in $2+\eps$ space-time dimensions \cite{Gastmans,Christensen:1978sc} or in the $1/N$ expansion \cite{Smolin:1981rm,Percacci:2006N}, a candidate for
such a NGFP has been recently observed in four dimensions thanks to the application of functional renormalization group equations (FRGE) to gravitational actions of the $f(R)$-type \cite{Souma,LR:2002,RS:2002,Litim:2008,OliverR2,CPR:2008,MS:2008,CPR2:2008}. Further support for AS comes  from lattice 
simulations \cite{Ambjorn:2005qt,AJL2}.

Since higher-derivative terms play a central role in the renormalizability and unitarity of gravity theories in the perturbative setting, it is of great interest to study their impact on the AS scenario. To this effect, in this letter
 we investigate the RG flow of gravity including all four-derivative interactions
and present, for the first time, non-perturbative results in this setting. In particular, our work augments the
previous $f(R)$ findings by including the second power-counting marginal coupling related to the Weyl-squared interaction.
This constitutes an important step in the program of studying the non-perturbative RG flow
of gravity in a derivative expansion, since, besides higher-derivative vertices studied previously, the
approximation also includes a four-derivative kinetic term for the graviton propagator.

Our main result is the existence of a NGFP with three UV-attractive and one UV-repulsive 
eigendirections. Together with the condition that this fixed point provides the continuum limit for gravity, the latter implies
a relation between the coupling constants of the theory at high energy. Tracing the RG flow to the IR, this translates into a relation between the four-derivative couplings in the effective field theory which could, in principle, be tested experimentally. Furthermore, the transition from asymptotic freedom to asymptotic safety may realize an old
proposal \cite{Julve:1978xn,Salam:1978fd} for the removal of the poltergeists from the spectrum of the theory, providing the exciting
perspective of asymptotically safe quantum gravity being unitary.

The key tool in our investigation is the effective average action $\Gamma_k$ and its FRGE \cite{Wetterich:1992yh} (see, e.g., \cite{Berges:2000ew,Niedermaier,Percacci:2007sz} for reviews). This setup provides a continuum analogue of Wilsons lattice renormalization group, realizing the idea of integrating out all fluctuation modes with momentum larger than a certain cutoff $k$ (high momentum modes), and taking them into account through a modified dynamics for the remaining fluctuations with momentum smaller than $k$ (low momentum modes). This is implemented by means of a cutoff $\cR_k$ which, for a given $k$, suppresses the contributions of the low momentum modes, so that only the high momentum modes are integrated out in the path integral. The functional $\Gamma_k$ then defines an effective field theory valid near the scale $k$ and, evaluated at tree level, describes all quantum effects originating from the high-momentum modes. In particular, $\Gamma_k$ essentially interpolates between the bare action $S_{\rm bare} \simeq
 \lim_{k \rightarrow \infty} \Gamma_k$ and the standard effective action $\Gamma = \lim_{k \rightarrow 0} \Gamma_k$. (For more details on the reconstruction of the bare action from $\Gamma_k$, see \cite{Manrique:2008zw}.)
The $k$-dependence of $\Gamma_k$ is governed by a FRGE which takes the form \cite{Reuter:1996cp}
\be\label{FRGE}
\p_t \Gamma_k[\Phi] = \half {\rm STr} \left[ \left( \frac{\delta^{2} \Gamma_k}{\delta \Phi \delta \Phi } + \cR_k \right)^{-1} \, \p_t \cR_k  \right]\, .
\ee 
Here, $t = \log(k/k_0)$, $\Phi$ represents all the fields in the theory, and $\rm STr$ denotes a generalized functional trace carrying a minus 
sign for fermionic fields and a factor 2 for complex fields.  
Lastly, the presence of the cutoff $\cR_k$ in the FRGE has the remarkable consequence of rendering the contribution of STr finite and peaked around $p^2 \approx k^2$; hence, an additional UV regularization of the trace is dispensable \cite{Niedermaier,Percacci:2007sz,Manrique:2008zw}.

As the FRGE \eqref{FRGE} cannot be solved exactly we have to resort to an approximation scheme.
A standard one is to take an ansatz (truncation) for $\Gamma_k[\Phi]$ of the form
\be\label{ansatz}
\Gamma_k[\Phi] = \sum_i u_{i}(k) \, \cO_i[\Phi]\,,
\ee
with a finite number of operators $\cO_i[\Phi]$ constructed from the fields $\Phi$, and with $u_i$ denoting scale-dependent coupling constants of mass dimension $d_i$. The  non-perturbative $\beta$-functions governing the RG flow of the $u_i$'s are obtained by substituting \eqref{ansatz} into \eqref{FRGE} and computing the coefficients of the operators $\cO_i[\Phi]$ appearing on the RHS. The reliability of the results so obtained is checked by systematically enlarging the truncation space.

In the following, we take $\Gamma_k[\Phi]$ to be of the form
\be\label{EAA}
\Gamma_k[\Phi] = \Gamma_k^{\rm gr} + S^{\rm gf} + S^{\rm gh} ,
\ee
where $\Gamma_k^{\rm gr}$, $S^{\rm gf}$, and $S^{\rm gh}$ are the gravitational part, 
the classical background-gauge fixing term, and the corresponding ghost action,  
respectively. The derivative expansion of $\Gamma_k^{\rm gr}$ up to fourth order is
given by
\be\label{eq:ansatz}
\Gamma_k^{\rm gr} = \int d^4x \sqrt{g} \left[ u_0 + u_1 R  - \frac{\omega}{3 \lambda} R^2 + \frac{1}{2\lambda} C^2 + \frac{\theta}{\lambda} E \right]\,,
\ee
where $R$ denotes the Ricci scalar, $C^2\equiv C_{\mu \nu \rho \sigma}C^{\mu \nu \rho \sigma}$ is the square of the Weyl tensor, and $E = R^2 - 4 R_{\mu \nu} R^{\mu \nu} + R_{\mu \nu \rho \sigma} \, R^{\mu \nu \rho \sigma}$ is the integrand of the topological Gauss-Bonnet term. 
The couplings $u_i$ are related to the dimensionful Newtons constant $G$ and the cosmological constant $\Lambda$ by
$u_0 = \L/(8 \pi G)$, $u_1 = - 1/(16 \pi G)$.
%
Truncations including the polynomials of the Ricci scalar have already 
been extensively analyzed in \cite{OliverR2,CPR:2008,MS:2008,CPR2:2008}. The inclusion of the tensor structures like $C^2$, 
responsible for the key features of higher-derivative gravity, is novel,
and provides an important test for the reliability of the approximations done so far in the literature. 
In particular, as explained in \cite{Benedetti:2009gn}, the inclusion of these terms is necessary in order to show that asymptotic safety is not spoiled by those same terms that make gravity perturbatively non-renormalizable.

The gauge fixing is carried out via the background field method.
In the presence of four-derivative operators, it is natural to consider a gauge-fixing term which also contains four derivatives.
 Following \cite{Codello:2006in} we use
\be
S^{\rm gf}[g;\gb] = \frac{\a}{2} \, \int d^4x \sqrt{\gb} \, F_\mu \, Y^{\mu \nu} \, F_\nu 
\ee
with $F_\mu = \Db^\nu h_{\mu \nu} - \frac{1}{4} \Db_\mu h $, $ Y^{\mu \nu} = \Db^2 \gb^{\mu \nu} $.
Here, $h_{\mu \nu} = g_{\mu \nu} - \gb_{\mu \nu}$ denotes the fluctuations of the metric $g_{\m\n}$ around the background $\gb_{\mu \nu}$, and $\Db_\mu$ 
the background-covariant derivative. 
Owing to the derivatives in $Y^{\mu\nu}$ the ghost sector of the theory contains, in addition to the usual complex ghost fields, a ``third" ghost \cite{Barth:1983hb}.
In the sequel, we will work in the limit $\a \rightarrow \infty$, which leads to considerable simplifications.

The construction of the non-perturbative $\beta$-functions for the couplings $u_i$ then proceeds by substituting the ansatz \eqref{EAA} 
into \eqref{FRGE} and projecting the RHS of the equation onto the curvature monomials contained in \eqref{eq:ansatz}. Going 
beyond the one-loop approximation \cite{Codello:2006in} thereby requires a class of background metrics which is generic enough to disentangle
 the coefficients multiplying $R^2$ and the tensorial terms, and, most importantly, simple enough to avoid the appearance of non-minimal 
higher-derivative differential operators inside the trace. While the maximally symmetric backgrounds used up to now are insufficient in 
the former respect, a generic compact Einstein background $\EE$ (which we take without Killing
 or conformal Killing vectors and without boundary for 
simplicity), satisfying
$\Rb_{\m\n}=\tfrac{\Rb}{4} \, \gb_{\m\n}$, 
is sufficient to meet both criteria and allows one to determine the non-perturbative $\beta$-functions of the linear combinations
\be
u_2 = - \frac{\omega}{3\lambda} + \frac{\theta}{6\lambda} 
\, , \qquad u_3 = \frac{1}{2\lambda} + \frac{\theta}{\lambda} \, .
\ee 
Finding a NGFP for these linear combinations implies that, barring miraculous cancellations, none of the couplings $\omega, \theta$ goes to infinity, and $\lambda$ and either $\omega$ or $\theta$ must be non-zero.

Substituting \eqref{EAA} and equating $g = \gb$ afterwards, the left-hand side of \eqref{FRGE} gives
\be \label{E-ansatz}
 \partial_t\bar{\Gamma}^{\rm gr}_k = \int d^4x \sqrt{\bar{g}} \big[  \dot{u}_0 + \dot{u}_1 \Rb +\left( \dot{u}_2 -\tfrac{1}{6} \dot{u}_3 
\right) \Rb^2 + \dot{u}_3 
\Rb_{\mu\nu\rho\sigma} \Rb^{\mu\nu\rho\sigma}  \big]\,.
\ee
To evaluate the RHS of \eqref{FRGE} we recast eq.\ \eqref{eq:ansatz} as
\be
\Gamma^{\rm gr}_k = \int  d^4x \sqrt{g} \big[  
u_0 + u_1 R + (u_2 - \tfrac{2}{3} u_3) R^2 
+ 2 u_3 R_{\m\n}R^{\m\n} + (u_3 + \tfrac{\theta}{\lambda}) E
\big] \,.
\ee
Noting that the variation of $E$ does not contribute to $\delta^2 \Gamma^{\rm gr}_k/\delta \Phi \delta \Phi$, the second variation can be 
obtained using the general results \cite{Barth:1983hb}, which considerably simplify once the background Einstein metric is substituted. 
Performing the transverse-traceless decomposition of the metric fluctuations and ghost fields \cite{York:1973ia} brings the operator inside the trace into (almost) diagonal form, with the off-diagonal terms vanishing in the limit $\alpha \rightarrow \infty$. Following \cite{CPR:2008,MS:2008,CPR2:2008}, the Jacobians arising from this change of variables can be represented 
as Gaussian integrals of appropriate auxiliary fields, in a similar fashion to the ghost 
representation of the Faddeev-Popov determinant.
The main virtue of this decomposition in combination with the choice of background is that all covariant derivatives and curvature tensors 
organize themselves in terms of Lichnerowicz Laplacians,
\be\label{eq:A4}
\Delta_{2L} \phi_{\mu\nu}  \equiv -\Db^2 \phi_{\mu\nu} - 2 \Rb_{\mu\,\,\,\nu}^{\,\,\,\alpha\,\,\,\beta} \phi_{\alpha\beta}\,,
\quad
\Delta_{1L} \phi_\mu   \equiv -\Db^2 \phi_\mu - \Rb_{\mu\nu} \phi^\nu\,, \; 
\quad \Delta_{0L} \phi  \equiv -\Db^2 \phi,
\ee
which then commute with all other quantities appearing under the traces.
Subsequently, the cutoff operators $\cR_k$ are constructed in such a way that the modified propagators are obtained by replacing
$z\to P_{n,k}(z)=z+\cR_{n,k}(z)$ for $z=\D_{nL}$. The resulting flow equation for the truncated $\Gamma_k$ then becomes
\be
\p_t \Gamma_k[\Phi] = \cS_{\rm 2T} + \cS_{\rm hh} + \cS_{\rm 1T} + \cS_{\rm 0} \, , 
\ee
where
\be
\begin{split}
\cS_{\rm 2T} = & \,  \tfrac{1}{2} \Tr_{\rm 2T} 
\left[
\tfrac{\p_t \left\{ 2 u_3 (P_{2,k}^2 - \D_{2L}^2)  - (u_1 + u_\flat R) R_{2,k} \right\}}
{2 u_3 P_{2,k}^2 - (u_1 + u_\flat R ) P_{2,k} - \tfrac{1}{2} u_1 R - u_0} 
\right], 
\quad \quad
\cS_{\rm 1T} =  \, - \tfrac{1}{2} \Tr_{\rm 1T} \left[ 
\tfrac{\p_t R_{1,k}}{P_{1,k}}
\right], \\ 
\cS_{hh} = &  \, \tfrac{1}{2} \Tr_{\rm 0} \left[
\tfrac{\p_t \left\{ 6 u_2 (P_{0,k}^2 - \D_{0L}^2) + (u_1 - 2  u_2 R) R_{0,k}  \right\}}{6 u_2 P_{0,k}^2 + (u_1 - 2 u_2 R) P_{0,k} + \tfrac{2}{3} u_0}
\right], 
\quad \quad
\cS_{\rm 0} =  \, - \tfrac{3}{2} \Tr_{\rm 0} \left[ 
\tfrac{ \p_t R_{0,k}}{3 P_{0,k} - R}
\right].
\end{split}
\ee
Here, $u_\flat = 2 u_2 - u_3/3$ and the indices ``2T'', ``1T'' and ``0'' indicate that the trace is over transverse-traceless matrices, transverse vectors and scalars, respectively.
These traces are then evaluated using general heat kernel methods 
adapted to the Lichnerowicz Laplacians. The resulting expressions are projected onto \eqref
{eq:ansatz} evaluated on an Einstein background, giving rise to the $\beta$-functions for the dimensionful couplings $u_i$.
 As a non-trivial check, we have verified explicitly that the expansion of the resulting $\beta$-functions to leading order in $\lambda$ precisely agrees with the (projected) universal one-loop result \cite{deBerredoPeixoto:2004if,Codello:2006in}.

In order to study the RG properties of the theory it is useful to switch to the dimensionless couplings $g_i = k^{-d_i} u_i$ and consider the $
\beta$-functions $\p_t g_i = \tilde{\beta}_i$. Our particular interest is on fixed points of the 
$\beta$-functions, $\tilde{\beta}_i(g_i) = 0$, which could provide a non-perturbative definition of quantum gravity within the AS program. 
We find that the $\beta$-functions for the couplings contained in \eqref{eq:ansatz} indeed
give rise to such a NGFP:\footnote{For conciseness we only give the numerical results for the optimized cutoff 
function $R_k(z)=(k^2-z)\th (k^2-z)$ \cite{Litim:2001up}, but we have verified that the use of a smooth exponential cutoff confirms the 
picture reported below.}
%
\be\label{NGFP}
 g_0^* = 0.00442 \; , \qquad g_1^* = -0.0101 \; ,
   \qquad g_2^* = 0.00754 \; , \qquad g_3^* = -0.0050 \, .
\ee
The ``universal'', i.e., gauge-independent \cite{deBerredoPeixoto:2004if}, combination $G \Lambda$ takes the fixed point value $(G\L)^*= 0.427$, which, together with the value of $g_2^*$,
%
%
is in good agreement with previous computations, notably the Einstein-Hilbert 
truncation \cite{Souma,LR:2002,RS:2002,Litim:2008}, the $R^2$-truncation \cite{OliverR2}, and the truncations using a polynomial expansion of $f(R)$-gravity \cite{CPR:2008,MS:2008,CPR2:2008}\footnote{Note that in previous works the background was assumed to be maximally symmetric, which means an Einstein space with zero Weyl tensor. In such case $R_{\m\n\r\s}R^{\m\n\r\s}=R^2/6$ and from \eqref{E-ansatz} it follows that one looses track of $u_3$. The value for $g_3^*$ 
constitutes a new result of asymptotically safe quantum gravity.} .
Notice that the finite values for $g_2^*$ and $g_3^*$ imply a finite value of $\l^*$, to be compared with the one-loop result $\l^*=0$. Thus the non-perturbative corrections captured by the FRGE shift the fixed point underlying the asymptotic freedom obtained within perturbation theory to the NGFP \eqref{NGFP} featuring in the asymptotic safety program.

An important characteristics of the NGFP are its stability properties.
The linearized RG flow around the NGFP, $\p_t g_i = {\bf B}_{ij} (g_j - g^*_j)$, is 
governed by the stability matrix ${\bf B}_{ij} \equiv \left. \p_j \beta_i \right|_{*}$ with stability coefficients (negative eigenvalues of $\bf B$)
\be\label{stab:coeff}
\theta_0 = 2.51  \, , \qquad  \theta_1 = 1.69 \, , \qquad
\theta_2 = 8.40  \, , \qquad  \theta_3 =  -2.11 \,  ,
\ee
and associated normalized eigenvectors
\be
\begin{split}
V_0 = & \, \{ \, 0.12 \, , \, 0.10 \,  , \,  -0.06 \, , \,   0.99  \}^{\rm T} \, ,
\quad \qquad V_1 =  \, \{ \, -0.20\, , \, 0.74 \,  ,  \, -0.10 \, , \,   0.63  \}^{\rm T} \, ,\\
V_2 = & \, \{ \, 0.74 \, , \, -0.46 \, , \, 0.48 \, , \, -0.11 \, \}^{\rm T} \, , 
\qquad V_3 =  \, \{ \, 0.07 \, , \, -0.21 \, , \, 0.97 \, , \, -0.09 \, \}^{\rm T} \, .
\end{split}
\ee
We observe that the inclusion of the $C^2$ coupling leads to real stability
coefficients at the NGFP. This is in contrast to the complex stability
coefficients and the corresponding spiraling approach of the RG flow
characteristic for $f(R)$-type truncations, and rather reflects the behavior
found within the perturbative one-loop computation \cite{Codello:2006in}.

Note that the transition from asymptotic freedom to asymptotic safety lifts the degeneracy of the marginal couplings. In particular, the negative stability coefficient $\theta_3$ indicates that the corresponding eigendirection is now UV-repulsive. The condition that a 
RG trajectory is asymptotically safe, i.e., that it approaches the fixed point as $k \rightarrow \infty$,
{\it implies a relation} between the couplings $g_i$. In the neighborhood of the NGFP, this relation allows to express $g_3$ in terms of the other couplings,
\be\label{critsurface}
g_3 = -0.116 + 0.745 g_0 - 2.441 g_1 + 11.06 g_2 \, .
\ee
It is remarkable that within our truncation the critical surface remains three-dimensional, as in the $f(R)$ case \cite{CPR:2008,MS:2008,CPR2:2008}.

\begin{figure}[h]
\vspace{-3mm}
\renewcommand{\baselinestretch}{1}
\epsfxsize=0.45\textwidth
\begin{center}
\leavevmode
\epsffile{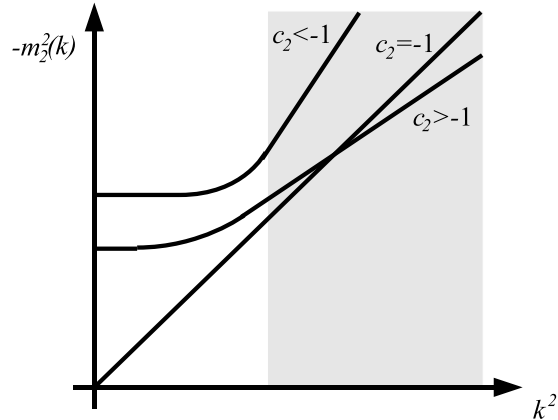} 
\end{center}
\caption{\label{fig1}{RG-scale dependence of the Poltergeist-mass. In the scaling regime of the NGFP (grey region) $m^2_2 \propto k^2$. For the proportionality factor $c_2 < -1$ the pole of the effective propagator is removed dynamically due to the running of the mass.}}
\end{figure}

A profound consequence of the UV limit of gravity being controlled by a NGFP is the possible removal of the poltergeist haunting higher derivative gravity, as proposed in \cite{Julve:1978xn,Salam:1978fd,Floreanini:1995}.
In this course, we observe that, 
besides the massless graviton and a massive scalar typical of $f(R)$ theories, classical actions of the form \eqref{eq:ansatz} give rise to a massive spin-2 poltergeist which is problematic for the unitarity of the resulting quantum theory (see, e.g., \cite{Stelle:1976gc,Stelle:1977ry}). 
Its mass is given by (minus) the pole of the (Euclidean) propagator, $m_{2}^{2}=\lambda/(16 \pi G)$.
In the quantum regime associated with the NGFP, the couplings $u_i$ depend on the RG scale,
%
$u_1 = k^2 g_1^*, u_2 = g_2^* \not = 0, u_3 = g_3^* \not = 0$. Thus, crucially, $\lambda^* \not = 0$,
%
and $m_2^2$ is not constant, but grows as $k^2$. This running naturally affects the position of the poles in the propagator. Given that $\Gamma_k$ provides an effective description for the physics at the scale $p^2 \approx k^2$, the mass has to be evaluated at this corresponding momentum scale. Identifying $p^2 = k^2$, the poltergeist propagator develops a pole at 
\be
k^2 = - m_{2}^{2}(k^2)\, .
\ee
There are then two possibilites. If  $\lambda^* > 0$, this effective mass diverges as $k \rightarrow \infty$, so that this unphysical mode
decouples from the theory in the UV. If $\lambda^* < 0$, then, provided the couplings enter the linear quantum regime where
$m_{2}^{2}(k^2)\sim c_2 k^2$ with $c_2 \equiv -g_1^* \lambda^*$, 
before the pole is met, the latter will be moved further by the RG running, and indeed completely removed if $c_2<-1$, viz. Fig.\ \ref{fig1}.
Unfortunately, our truncation is not precise enough to determine $\lambda^*$ and therefore draw any definite conclusions.
Note, however, that the general mechanism here does not rely on our truncation details, but is a generic consequence of the NGFP. The fact the theory is asymptotically safe instead of asymptotically free is vital for this argument.

To summarize, the $\beta$-functions of our higher-derivative truncation possess a non-trivial fixed point with three UV-stable and one UV-repulsive eigendirections, in accordance with the asymptotic safety scenario. While a proof of asymptotic safety for gravity would, most likely, include the analysis of an infinite series of curvature terms, which is  beyond current FRG-methods, the mutually consistent results obtained by systematically extending the truncation subspaces under consideration lends strong support to the idea that 
gravity is indeed asymptotically safe.
Moreover, the transition from asymptotic freedom to asymptotic safety provides an interesting new perspective on the ``poltergeists'' appearing in higher-derivative gravity theories, so that asymptotically safe quantum gravity may actually be safe from such hauntings.

\section*{Acknowledgements}

We thank A.\ Codello, R.\ Loll, R.\ Percacci and M.\ Reuter for discussions.   
Research at Perimeter Institute is supported in part by the Government of Canada through NSERC and by the Province of Ontario through 
MRI. P.F.M. is supported by the Netherlands Organization for Scientific Research (NWO) under their VICI program. F.S.\ acknowledges 
financial support from the ANR grant BLAN06-3-137168.



\begin{thebibliography}{10}

\bibitem{Stelle:1976gc}
K.~S. Stelle,
{\it Phys. Rev. }{\bf D16}, 953 (1977).

\bibitem{Julve:1978xn}
J.~Julve and M.~Tonin,
{\it Nuovo Cim. }{\bf B46}, 137 (1978).

\bibitem{Fradkin:1981hx}
E.~S. Fradkin and A.~A. Tseytlin,
{\it Phys. Lett. }{\bf B104}, 377 (1981);
{\it Nucl. Phys. }{\bf B201}, 469 (1982).


\bibitem{Avramidi:1985ki}
I.~G. Avramidi and A.~O. Barvinsky,
{\it Phys. Lett. }{\bf B159}, 269 (1985).

\bibitem{deBerredoPeixoto:2004if}
G.~de~Berredo-Peixoto and I.~L. Shapiro,
{\it Phys. Rev. }{\bf D71}, 064005 (2005), hep-th/0412249.

\bibitem{Codello:2006in}
A.~Codello and R.~Percacci,
{\it Phys. Rev. Lett. }{\bf 97}, 221301 (2006), hep-th/0607128.

\bibitem{Stelle:1977ry}
K.~S. Stelle,
{\it Gen. Rel. Grav. }{\bf 9}, 353 (1978).

\bibitem{Salam:1978fd}
 A.~Salam and J.~A.~Strathdee,
{\it Phys.\ Rev.\  D }{\bf 18}, 4480 (1978).

\bibitem{Weinberg:1980gg}
S.~Weinberg,
in {\it General Relativity}, eds.~ S.W. Hawking, W. Israel (Cambridge Univ.\
  Pr. 1980).

\bibitem{Niedermaier}
 M.~Niedermaier and M.~Reuter,
{\it  Living Rev.\ Rel.\  }{\bf 9}, 5 (2006).

\bibitem{Percacci:2007sz}
R.~Percacci,
arXiv:0709.3851.

\bibitem{Reuter:2007rv}
M.~Reuter and F.~Saueressig,
{\it arXiv:0708.1317}.

\bibitem{Gastmans}
R.~Gastmans, R.~Kallosh, and C.~Truffin,
{\it Nucl. Phys. }{\bf B133}, 417 (1978);
\bibitem{Christensen:1978sc}
S.~M. Christensen and M.~J. Duff,
{\it Phys. Lett. }{\bf B79}, 213 (1978).

\bibitem{Smolin:1981rm}
L.~Smolin,
{\it Nucl. Phys. }{\bf B208}, 439 (1982).

\bibitem{Percacci:2006N}
 R.~Percacci, {\it Phys.\ Rev.\ D }{\bf 73}, 041501 (2006), hep-th/0511177.

\bibitem{Souma}
W.~Souma,
{\it  Prog.\ Theor.\ Phys.\  }{\bf 102}, 181 (1999), 
  hep-th/9907027.
  
\bibitem{LR:2002}
O.~Lauscher and M.~Reuter,
{\it  Phys.\ Rev.\  D }{\bf 65}, 025013 (2002),
  hep-th/0108040.
  
\bibitem{RS:2002}
M.~Reuter and F.~Saueressig,
{\it  Phys.\ Rev.\  D }{\bf 65}, 065016 (2002), hep-th/0110054.

\bibitem{Litim:2008}
D.~F.~Litim,
arXiv:0810.3675.



\bibitem{OliverR2}
O.~Lauscher and M.~Reuter,
{\it Class. Quant. Grav. }{\bf 19}, 483 (2002), hep-th/0110021;
{\it Int. J. Mod. Phys. }{\bf A17}, 993 (2002), hep-th/0112089;
{\it Phys. Rev. }{\bf D66}, 025026 (2002), hep-th/0205062.


\bibitem{CPR:2008}
A.~Codello, R.~Percacci, and C.~Rahmede,
{\it Int. J. Mod. Phys. }{\bf A23}, 143 (2008), arXiv:0705.1769.

\bibitem{MS:2008}
P.~F.~Machado and F.~Saueressig,
{\it Phys.\ Rev.\   }{\bf D77}, 124045 (2008),
arXiv:0712.0445;

\bibitem{CPR2:2008}
A.~Codello, R.~Percacci, and C.~Rahmede,
arXiv:0805.2909.

\bibitem{Ambjorn:2005qt}
J.~Ambj{\o}rn, J.~Jurkiewicz, and R.~Loll,
{\it Phys.\ Rev.\ Lett.\ }{\bf 95}, 171301 (2005), hep-th/0505113;
%
{\it Phys. Rev. }{\bf D72}, 064014 (2005), hep-th/0505154.

\bibitem{AJL2}
J.~Ambj{\o}rn, A.~G{\"o}rlich, J.~Jurkiewicz, and R.~Loll,
{\it Phys. Rev. }{\bf D78}, 063544 (2008), arXiv:0807.4481.

\bibitem{Floreanini:1995}
R.~Floreanini and R.~Percacci,
{\it Phys. Rev. }{\bf D52}, 896 (1995), hep-th/9412181 .

\bibitem{Berges:2000ew}
  J.~Berges, N.~Tetradis and C.~Wetterich,
{\it  Phys.\ Rept.\  }{\bf 363} (2002) 223, 
  hep-ph/0005122.



\bibitem{Reuter:1996cp}
M.~Reuter,
{\it Phys. Rev. }{\bf D57}, 971 (1998), hep-th/9605030.

\bibitem{Wetterich:1992yh}
C.~Wetterich,
{\it Phys. Lett. }{\bf B301}, 90 (1993).

\bibitem{Manrique:2008zw}
  E.~Manrique and M.~Reuter,
  {\it Phys.\ Rev.\  D }{\bf 79} (2009) 025008
  [arXiv:0811.3888 [hep-th]].

\bibitem{Benedetti:2009gn}
  D.~Benedetti, P.~F.~Machado and F.~Saueressig,
  arXiv:0902.4630 [hep-th].


\bibitem{Barth:1983hb}
N.~H. Barth and S.~M. Christensen,
{\it Phys. Rev. }{\bf D28}, 1876 (1983).

\bibitem{York:1973ia}
J.~W.~York,
{\it  J.\ Math.\ Phys.\  }{\bf 14} (1973) 456.

\bibitem{Litim:2001up}
D.~F. Litim,
{\it Phys. Rev. }{\bf D64}, 105007 (2001), hep-th/0103195.



\end{thebibliography}
\end{document}